\documentclass[aps,showpacs,onecolumn,eqsecnum,floatfix]{revtex4}

\usepackage{epsfig}\usepackage{latexsym}

\begin{document}


\title{A unified Scenario for modeling the Galactic and Cosmological Dark 
Matter Components}

\author{Jos\'e A. Gonz\'alez}
\email{cervera@nuclecu.unam.mx}

\author{Daniel Sudarsky}
\email{sudarsky@nuclecu.unam.mx}

\affiliation{Instituto de Ciencias Nucleares, Universidad Nacional
Aut\'onoma de M\'exico, A.P. 70-543, M\'exico D.F. 04510, M\'exico.}


\date{\today}


\begin{abstract}In this work we analyze the viability of use a particular models of scalar
fields in the context of the galactic dark matter problem.These models are based on a single 
scalar field, minimally coupled to the gravity in a asymptotically flat or asymptotically de 
Sitter spacetime. We discuss the opening possibility of constructing a unified model for both
the cosmological and the galactic dark matter.
\end{abstract}


\pacs{
04.20.-q   
04.70.Bw   
95.35.+d   
}


\maketitle


\section{Introduction}
\label{sec:introduction}

The problem of the galactic dark matter lies in the fact that the rotational curves of the 
spiral galaxies (away from the galactic center) are flat while the expected behavior is a 
keplerian decay, and that the total gravitational pull inferred, indicates an amount of gravitating  
matter which exceeds by almost an order of magnitude the luminous matter within the galaxy.
A second dark matter problem is the cosmological dark matter problem, the fact that the data about 
the cosmological evolution seems to require the mean density of the universe to be quite close to 
the critical density while ordinary matter can only account for a small fraction 
thereof \cite{Nucleosynthesis}. The evidence in this regard comes from recent observation of distant 
type Ia supernovas \cite{Ia}, and  from  the latest data about the CMB anisotropies \cite{Wmap},
all of which is in accordance with the standard inflationary prediction that $\Omega_{total}=1$ \cite{Guth}.
The latter problem has been considered as a completely separated issue and has lead even to its  
remaining as the {\it dark energy problem}. The approach to this has centered in the introduction of a 
new type of scalar field dubbed the``Quintessence". 

We will briefly consider the possibility of connecting the two problems through scenarios that hope to 
treat both simultaneously. Regarding the galactic dark matter problem two kinds of models are usually
considered: The first one consists in exotic dark matter in the form of very weakly interacting particles, 
while the second one consists in classical field configurations. The former enjoy a widespread popularity, 
specially amongst astronomers who are used to regard particles as the natural constituents of almost all 
kinds of matter, while many physicist are more attracted by the second type due to the fact that the 
particle concept is regarded as secondary, and available only under certain conditions \cite{Wald}. 
However we know that both the classical field and the particle models have a fundamental underlying
quantum field model, and whether the field or the particle descriptions are the most appropriate depends 
on the circumstances. In fact particles and classical fields represent two distinct classical limits of
a quantum field.

It is well known that the two descriptions are rather analogous to the momentum and position descriptions 
of elementary quantum mechanics as the two relevant operators do not commute. In fact if we consider a scalar 
field $\hat \phi(x) = \sum_s \left\{ a_s e^{-i k \cdot x}+ a^{\dag}_s e^{i k \cdot x}\right\}$ 
and $ \hat N = \sum_r a^{\dag}_r a_r$ the total number operator for the corresponding particles, we easily 
see that they do not commute: 
$\lbrack \hat N, \hat \phi(x) \rbrack = \sum_r \left\{ a^{\dag}_r e^{i k \cdot x}- a_r e^{-i k \cdot x} \right\} \ne 0 $.
Thus when considering the first types of models, astronomers have to introduce not only the explicit 
--and not quite self evident-- hypothesis regarding the evolution of the gas of dark matter particles,
but also, implicitly a series of assumptions that are often inadvertently made regarding the mechanisms 
by which the dark matter field underlying the dark mater came to be in the type of incoherent configuration 
that justifies the particle description. 

Regarding the second type of models, we must point out that a major difficulty that has to be overcame is the 
increasing evidence regarding the presence, in most galaxies, of large central black holes. A model based on 
fields must therefore, be able to represent a static stable configuration of the fields, with no sources 
(to avoid even more assumptions about additional new types of matter that would represent such sources,
given the fact that there are very stringent limits --associated with test of the equivalence principle-- 
on ordinary matter couplings to long range fields \cite{equiv_princ}) and it must admit black holes at the 
center of the galaxies in order to be a viable model of the galactic dark matter. 

In the single scalar field arena, it had not been considered possible to obtain models with black holes and 
static stable configurations of scalar fields, due the no-hair theorems that ensure the no existence of such
configurations \cite{No Hair}. Even the so called boson star configurations cease to exits when black holes 
are introduced \cite{Pena}.

Recently, however scalar hairy solutions in asymptotically Anti de-Sitter spacetimes have been found \cite{Torii}. 
In a recent analysis \cite{Gonzalez} we proposed the existence of solutions of this kind  in asymptotically flat
regime, which where latter found in \cite{Nucamendi}. Such solutions escape the no-hair theorems due the fact 
the energy conditions considered in the theorems do not hold in the corresponding models.
In this essay we consider the question of whether such type of models could be a good candidate for the unified 
treatment of the clustered dark matter and the cosmological dark energy problems. At this point we should 
mention previous attempts to construct such unifying scenarios, which are generically dubbed ``Quartessence''
models \cite {Waga}. The first kind is based in a purely phenomenological fluid model known as the 
``Chaplygin Gas'' which is capable of exhibiting under different conditions the types of equations of state 
associated with both the dark energy and dark matter. The second approach exemplified by the works \cite{Tachyon} 
deals with a string inspired tachyonic field, again capable of exhibiting the two different types of equations 
of state required, but which as far as we know, have not dealt at all with the serious issue that a black hole 
at the galactic centers would posse for the existence of standard scalar field configurations as mentioned above.

In \ref{sec:model} we describe the model and in \ref{sec:discussion} we discuss it's viability as well as some 
of the aspects that need to be further explored.


\section{The model}
\label{sec:model}  

\begin{figure}
\vspace{5mm}
\epsfig{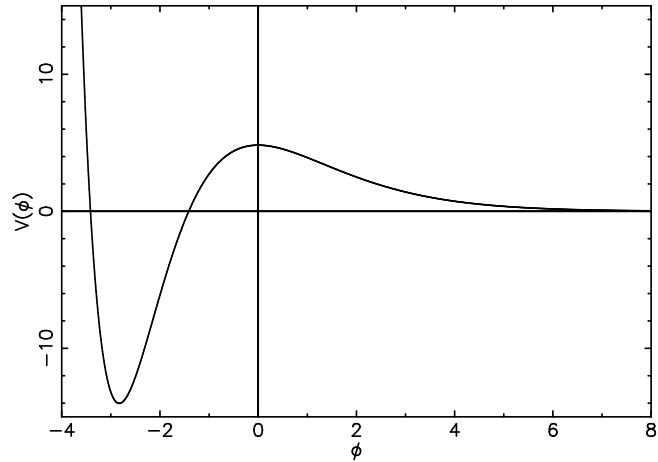}
\caption{Potential \ref{potential} with $V_0 = a = \beta = 1$ and $s = -\frac{1+\sqrt{1+a\beta^2}}{\beta}$.}
\label{fig:potential}
\end{figure}

We will consider a model of a scalar field minimally coupled to gravity and with a potential having the generic 
features shown in Fig. \ref{fig:potential}. The model is thus based on the following Lagrangian:

\begin{equation}
{\cal L} = \sqrt{-g} \left[ { 1\over 16\pi } R- {1\over 2}\nabla_{\alpha} \phi \nabla^{\alpha} \phi- V(\phi) \right]  \; . 
\label{lag}
\end{equation}

An example of the kind of potential requires is given by:

\begin{equation}
\label{potential}
V(\phi) = V_0 ((\phi-s)^2-a) e^{- \beta \phi} \; ,
\end{equation}

where $V_0$, $a$, $s$ and $\beta$ are constants.

The gravitational and scalar field equations following from 
the Lagrangian (\ref{lag}) can be written as

\begin{equation}
G_{\mu\nu} = 8\pi T_{\mu \nu} \,\,\,,\,\,\,\Box \phi 
= \frac{\partial V(\phi)}{\partial\phi}\,\,\,\,,
\end{equation}

where

\begin{equation}
T_{\mu\nu} = \nabla_\mu \phi \nabla_\nu \phi 
- g_{\mu\nu}\left[\frac{1}{2} \nabla_{\alpha} \phi \nabla^{\alpha} \phi
+ V(\phi) \right]
\end{equation}  

When considering the galactic dark matter we focus on the static spherically symmetric metric 
parametrized as

\begin{equation}
ds^2= -e^{2 \delta} \mu dt^2 + \mu^{-1} dr^2 + r^2 d\Omega^2 \; ,
\end{equation}

Where $\mu(r) = 1 - \frac{2 m(r)}{r}$.  
Thus Einstein's equations take the form:

\begin{equation}
\partial_r m = 4 \pi r^2 \left[ \frac{\mu (\partial_r \phi)^2}{2} + V(\phi)\right] \; , \qquad
\partial_r \delta = 4 \pi r (\partial_r \phi)^2 \; ,
\end{equation}

and the scalar field equation can be written as

\begin{equation}\partial_{r r}^2 \phi = - \left[\frac{2}{r} + \frac{2}{\mu} \left\{ \frac{m}{r^2}
- 4 \pi r V(\phi)\right\} \right] \partial_r \phi+ \frac{1}{\mu}\frac{\partial V(\phi)}{\partial \phi} \; ,
\end{equation}

this type of potentials are known to lead, as all ready  mentioned to non trivial field configurations 
containing central black holes in the asymptotically flat and asymptotically Anti de Sitter case and 
our heuristic analysis indicates that the solutions should also exist in the asymptotically de Sitter 
case (when the asymptotic value of the potential is positive). Moreover a similar solution can be 
expected  when the scalar field at infinity is not exactly in the false vacuum but is close enough,  
including a situation where in the asymptotic region it is slowly rolling down towards that value!.

In the cosmological case one focus on a $k=0$ Robertson Walker 4 spacetime

\begin{equation}
ds^2= - dt^2 + a^2 ( dr^2 + r^2 d\Omega^2 ) \; ,
\end{equation}  

that Einstein's equations take the well known  form

\begin{equation}
\left(\frac{\partial_t a}{a}\right)^2 = \frac{4\pi}{3} \left[ (\partial_t \phi)^2 + 2 V(\phi) \right] 
\qquad \qquad ,\frac{ \partial_t^2 a}{a} = - \frac{8\pi}{3}[(\partial_t \phi)^2 - V(\phi) ] \;. 
\end{equation}

Thus in this regard we could be essentially in the so called quintessence models based on a scalar field  
slowly rolling down to a minima of a potential (the one corresponding to  $\phi = +\infty$) .  
Recall that in these schemes the scalar field is spatially homogeneous and it is evolving in time, 
rolling down ever slowly under the combined influence of the cosmic expansion and its self interaction 
potential because the potential can considered as an exponentially decaying one such as 
$V_{Quintessence}= V_0 e^{-\phi}$.  
The {\it unifying} scenario we are proposing is then the following: 
At early enough times in cosmic history, when the universe was hot enough, the scalar field would be in 
its false vacuum $\phi=0$ (as in the scenarios of spontaneous symmetry breaking) and as the universe cools 
the scalar field will start to move to the minima of the potential. In those regions where the field started 
to move towards the negative values we would have formation of solitons and eventually black holes \cite{Alcubierre}
that would account for the early structure formation leading to the attraction of barionic matter after its 
decoupling from radiation, and to the present day galaxies, while those regions where the field started to 
move towards the positive values would evolve into the intergalactic regions where the cosmic expansion of the 
quintessence type would take place. The equality in the order of magnitude of the contributions to the  
total energy density of the universe in the  form of clumped ``dark matter", and the essentially homogeneously 
distributed ``dark energy", would be a result of the approximate symmetry between positive and negative valued 
fluctuations of the scalar field about its initial mean value $\phi =0$.


\section{Discussion}
\label{sec:discussion}

We have considered the possible use of single scalar field model for the galactic dark matter which are consistent 
with the existence of large black holes in the center of the galaxies, as well as the cosmological dark energy components 
of the universe. The reluctance of considering field models in the first context should be tempered by the observation 
that magnetic fields are unquestionable components (of relative small quantitative value) of the galactic matter content, 
and the only fundamental reason behind the difference in the way they are considered seems to be related to familiarity,
a criteria that can hardly be justified when dealing with the nature of the dark matter. 
We have argued that in fact the field models should be considered superior from the point of view of number of hypothesis  
involved (in the spirit of Occam's razor), to the standard models base on gases of very weakly interacting particles, 
due to the large number of unmentioned underlying assumptions that are involved in the latter. We have seen that single 
scalar field models open the possibility of regarding the dark matter (the clumped component) and the dark energy
(the cosmological component) within a unified context. Most astronomers tend to regard this two problem as completely 
different issues, and even to apply very different criteria of naturalness when considering the two. However we must 
also point out the estimates of the contribution of both types of matter to the total cosmic mean density are of the 
same order of magnitude, making a unified approach all the more desirable.
Moreover we would have at the same time a new approach to understanding the origin of the super massive
black hole at the galactic centers. There are of course further issues to 
investigate, such as the precise form of the potential of the model that would account for the different rotation curves 
in the different galaxies, the correlations between the rotation curves and the central black holes \cite{correlations},
(we must mention that the introduction of a non minimal coupling for the scalar field holds some promise regarding 
these last two issues as has been argued in a study of such questions in a different approach \cite{NonMinimal})
the interaction between the configurations associated with different galaxies, the behavior of these solutions under 
the effects general cosmic expansion, and in fact their cosmic evolution starting from the primordial density perturbations.
In many of these regards the scenario looks very promising at this very early point, however it is clear that all of 
them should be the foccus of further research.



\end{document}